# Deformation behavior of Mg-8.5wt.%Al alloy under reverse loading investigated by *in-situ* neutron diffraction and elastic viscoplastic self-consistent modeling

H. Wang [a, *], S.Y. Lee [b,†], M.A. Gharghouri [c], P.D. Wu [d]

[a] Materials Science and Technology, Los Alamos National Laboratory, Los Alamos, NM, USA, 87544
[b] Department of Materials Science and Engineering, Chungnam National University, Daejeon 305-764, South Korea
[c] Canadian Neutron Beam Centre, Canadian Nuclear Laboratories, Chalk River, ON K0J 1J0, Canada
[d] Department of Mechanical Engineering, McMaster University, Hamilton, ON, L8S 4L7, Canada

**Abstract**

The cyclic deformation behavior of extruded Mg-8.5wt.%Al alloy with a conventional extrusion texture and a modified texture is systematically investigated by *in-situ* neutron diffraction and elastic viscoplastic self-consistent (EVPSC) modeling incorporating a twinning/de-twinning (TDT) scheme. The role of twinning and de-twinning on the deformation behavior of Mg-8.5wt.% Al alloy is investigated in terms of the macroscopic stress-strain response, the evolution of the activities of various deformation mechanisms, the texture evolution, the evolution of the internal elastic strains, and the evolution of the diffraction peak intensities. The alloy with the conventional extrusion texture undergoes twinning during initial compression and de-twinning during reverse tension. The same alloy does not favor twinning during initial tension, but rather during reverse compression. The alloy with a modified texture undergoes twinning during initial tension followed by detwinning during reverse compression. The results provide insights into the effect of initial texture, loading path, slip, twinning, de-twinning on the cyclic behavior of magnesium.

[*] Corresponding author. Dr. Huamiao Wang, Tel.: +505-664-0321, E-mail: wanghm@lanl.gov.
[†] Corresponding author. Dr. Soo Yeol Lee, Tel.: +82-42-821-6637, Fax: +82-42-822-5850, E-mail: sylee2012@cnu.ac.kr

# 1. Introduction

Magnesium alloys have been intensively studied in the past few decades because they are the lightest structural materials and therefore offer the potential to reduce energy consumption in the transportation sector through light-weighting [1]. The application of conventional (non Rare-Earth containing) magnesium alloys is inhibited by their poor formability at room temperature, which is due to the limited number of active deformation mechanisms. Conventional wrought magnesium alloys show strongly anisotropic behavior due to the development of crystallographic texture during processing, in which the c-axis is preferentially oriented perpendicular to the processing direction (rolling or extrusion direction) [2-6].

Slip and twinning can both contribute significantly to plastic deformation in conventional magnesium alloys. Unlike slip, deformation twining can only be activated unidirectionally. As a result, twinning activity depends strongly on the orientation of the crystal lattice relative to the applied loading direction. The twinned region is reoriented to mirror the parent lattice orientation. Extension twinning is commonly observed in magnesium alloys at room temperature, resulting in a lattice reorientation of 86.3°. Upon load reversal, the twinned regions can undergo de-twinning [6-10]. Both twinning and detwinning have been observed during cyclic deformation and loading path changes [5].

The macroscopic behavior of a polycrystalline material is a function of the response of each constituent grain, which in turn depends on the orientation of the grain with respect to the loading axes as well its interactions with the surrounding grains. Therefore, the grain level behavior is of great importance. In-situ neutron diffraction experiments provide quantitative information on the grain-level behavior; the technique has been applied to many materials (e.g., stainless steel, Mg alloys, Zircaloy, etc. [2,3,5,11-14].

During cyclic deformation, magnesium alloys are known to undergo twinning and de-twinning [5-9]. However little information is available on the evolution of the internal elastic strain during cyclic deformation of magnesium alloys [15-18]. By combining *in-situ* measurements with modeling, it is possible to obtain detailed quantitative information about the nature and extent of slip and twinning. The Elastic Viscoplastic Self-Consistent (EVPSC) model of polycrystal plasticity has been widely applied to study the macroscopic behavior [19-26] and internal elastic strain evolution of magnesium alloys under monotonic loading [27,28]. The predominant twin reorientation (PTR) scheme has been used to treat twinning but, while it accounts for twinning, it does not account for detwinning. The Twinning / De-Twinning (TDT) scheme was thus developed and implemented in the EVPSC model to account for the effects of both twinning and detwinning (EVPSC-TDT) [29-35]. To date, the EVPSC-TDT model has been used to model the behavior of ZK60A extruded bar under symmetric low-strain cyclic straining [16], and the behavior of AZ31B rolled plate under asymmetric cyclic straining [17]. In these two studies, the deformation texture was such that deformation twinning was activated mainly by compression perpendicular to the c-axis of the HCP lattice. In contrast, here, for the first time, we apply the EVPSC-TDT model to investigate a situation in which twinning is activated by tension along the c-axis of the HCP lattice.

Taking advantage of both *in-situ* neutron diffraction and EVPSC-TDT modelling, the behavior of Mg-8.5wt.%Al alloy under cyclic deformation is systematically investigated in the current work. A sample with a conventional extrusion texture (c-axes perpendicular to the extrusion direction (ED)) and a sample with a modified texture (c-axes parallel to ED) are deformed cyclically (tension followed by reverse compression and compression followed by reverse tension). Both the macroscopic behavior (stress-strain relation and texture evolution) and the microscopic behavior (internal elastic strains and the diffraction peak intensities) are experimentally examined and numerically simulated.

## 2. Experimental procedure

An extruded wrought Mg-8.5wt.%Al alloy was prepared at the Péchiney Research Centre, France. The material was solution treated and aged. A detailed description of the sample preparation is provided elsewhere [2]. Bulk crystallographic texture measurements were performed using the E3 neutron diffractometer of the Canadian Neutron Beam Centre, located in the NRU reactor, Chalk River Laboratories, Canada. The orientation distribution function (ODF) for each sample was determined from four pole figures ({10.0}, {00.2}, {10.1} and {10.2}). Two starting textures were used: (1) as-extruded texture, T1 (Fig. 1a), in which the basal poles of most grains are oriented normal to the extrusion axis and a small portion of grains are oriented with the basal pole parallel to the extrusion axis; (2) modified texture, T2 (Fig. 1b), in which the basal poles of most grains are oriented parallel to the extrusion axis. Thus, {10.2} extension twinning could be easily activated under compression and tension along the extrusion direction for T1 and T2, respectively.

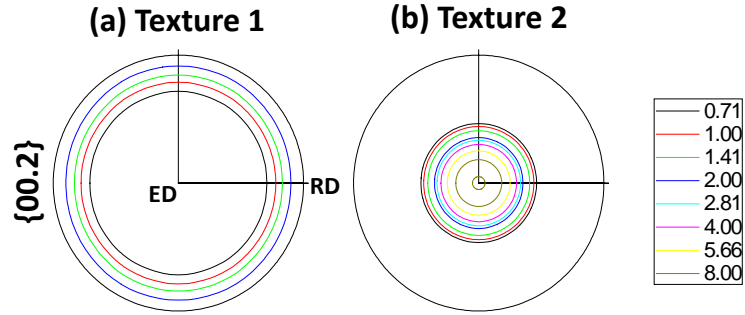

Figure 1. Initial textures of (a) T1 and (b) T2 samples determined by neutron diffraction. Note that the center of each pole figure corresponds to the extrusion direction (ED), and RD is the radial direction of the extruded bar.

*In-situ* neutron diffraction experiments were performed on the L3 neutron diffractometer of the Canadian Neutron Beam Centre. Three *in-situ* neutron diffraction tests were conducted: 1) T1 sample deformed in Tension followed by reverse Compression (denoted T1-TC); 2) T1 sample deformed in Compression followed by reverse Tension (T1-CT); and 3) T2 sample deformed in Tension followed by reverse Compression (T2-TC). Diffraction peaks for several grain orientations, with the plane normal parallel to the loading direction, were measured *in-situ* during deformation. The interplanar spacings (*d*-spacings) of various crystallographic orientations along the axial direction (parallel to the extrusion direction (ED)) were measured in terms of the applied load. The internal elastic strain for a given plane {*hk.l*} is given by:

$$\varepsilon_L^{hk.l} = \frac{d^{hk.l} - d_0^{hk.l}}{d_0^{hk.l}} \quad (1)$$

where $d_0^{hk.l}$ and $d^{hk.l}$ are the *d*-spacings corresponding to the un-deformed (reference) and deformed states, respectively [36,37]. Changes in diffraction peak intensity over the small strain intervals used here are due mainly to twinning and detwinning. The internal elastic strains and intensities allow us to examine the onset and propagation of twinning/de-twinning and the associated stress state in the un-twinned regions and the twinned regions at the grain level. Part of the experimental results was reported previously [18].

## 3. EVPSC-TDT model

A brief description of the TDT scheme is provided below. A more detailed description of the TDT scheme and its implementation in the EVPSC model is provided elsewhere [29,30]. The elastic deformation of a crystal is related to its stress state through the elastic constants, while the plastic deformation is accommodated by slip and twinning on systems ($\boldsymbol{s}^\alpha, \boldsymbol{n}^\alpha$). Here, $\boldsymbol{s}^\alpha$ and $\boldsymbol{n}^\alpha$ are the slip/twinning direction and the direction normal to the slip/twinning plane for system $\alpha$, respectively. The plastic strain rate tensor for a crystal can be expressed as:

$$\dot{\boldsymbol{\varepsilon}}^p = \sum_\alpha \dot{\gamma}^\alpha \boldsymbol{P}^\alpha \qquad (2)$$

in terms of the shear rate $\dot{\gamma}^\alpha$ and the Schmid tensor $\boldsymbol{P}^\alpha = (\boldsymbol{s}^\alpha \boldsymbol{n}^\alpha + \boldsymbol{n}^\alpha \boldsymbol{s}^\alpha)/2$ for system $\alpha$. For both slip and twinning, the shear rate $\dot{\gamma}^\alpha$ is related to the resolved shear stress $\tau^\alpha = \boldsymbol{s}^\alpha \cdot \boldsymbol{\sigma} \cdot \boldsymbol{n}^\alpha = \boldsymbol{\sigma}:\boldsymbol{P}^\alpha$, where $\boldsymbol{\sigma}$ is the Cauchy stress tensor.

For slip, the shear rate on system $\alpha$ can be described by the following relation:

$$\dot{\gamma}^\alpha = \dot{\gamma}_0 |\tau^\alpha/\tau_{cr}^\alpha|^{1/m} \mathrm{sgn}(\tau^\alpha) \qquad (3)$$

where $\dot{\gamma}_0$ is a reference shear rate, $\tau_{cr}^\alpha$ is the critical resolved shear stress (CRSS), and $m$ is the strain rate sensitivity. The evolution of the critical resolved shear stress (CRSS) $\tau_{cr}^\alpha$ is given by:

$$\dot{\tau}_{cr}^\alpha = \frac{d\hat{\tau}^\alpha}{d\Gamma} \sum_\beta h^{\alpha\beta} |\dot{\gamma}^\beta| \qquad (4)$$

where $\Gamma = \sum_\alpha \int |\dot{\gamma}^\alpha| dt$ is the accumulated shear strain in the grain, and $h^{\alpha\beta}$ are the latent hardening coupling coefficients which empirically account for the obstacles on system $\alpha$ associated with system $\beta$. $\hat{\tau}^\alpha$ is the threshold stress and is defined by an extended Voce law:

$$\hat{\tau}^\alpha = \tau_0^\alpha + (\tau_1^\alpha + h_1^\alpha \Gamma)\left(1 - \exp\left(-\frac{h_0^\alpha}{\tau_1^\alpha}\Gamma\right)\right) \qquad (5)$$

where $\tau_0$, $h_0$, $h_1$ and $\tau_0 + \tau_1$ are the initial CRSS, the initial hardening rate, the asymptotic hardening rate, and the back-extrapolated CRSS, respectively. Equations (3-5) govern slip activation and hardening, but do not apply to twinning, which must be treated differently to account for its polar nature, and for the associated lattice reorientation.

In the TDT scheme, twin nucleation within an existing twin-free grain is first simulated by creating a new grain whose orientation is related to that of the original grain via the twin relationship. The untwinned section of the original grain is the *matrix* while the twinned region is the *twin*. The volume of the newly-created twin is taken to be 0.1% of the original untwinned grain volume, and the volume of the matrix is simultaneously decreased by the same amount. Furthermore, the stress and hardening of the twin are assumed to be the same as those of the matrix. The twin can grow (twinning) or shrink (de-twinning), but the total volume fraction of the twin and matrix does not change. As deformation proceeds, the twin and matrix are allowed to evolve as independent grains in the Homogeneous Effective Medium, except that their total combined volume fraction remains constant.

Twin growth is accomplished via matrix reduction (MR) and twin propagation (TP), while twin shrinkage is accomplished via matrix propagation (MP) and twin reduction (TR). MR and MP are driven by the stress in the matrix, while TR and TP are driven by the stress in the twin. Furthermore, the plastic strain MR and MP is attributed to the matrix, while the plastic strain due to TP and TR is attributed to the twin. For MR and TR, the shear rate associated with twinning system $\alpha$ can be written as:

$$\dot{\gamma}_I^\alpha = \begin{cases} \dot{\gamma}_0 |\tau^\alpha/\tau_{cr}^\alpha|^{1/m} & \tau^\alpha > 0 \\ 0 & \tau^\alpha \leq 0 \end{cases} \quad (6)$$

where the subscript $I$ represents MR or TR. For MR, the resolved shear stress, $\tau^\alpha$, is calculated from the stress in the matrix, with $\tau_{cr}^\alpha$ the critical resolved shear stress for propagation of the twin on twin system $\alpha$. For TR, $\tau^\alpha$ is calculated from the stress in the twin, with $\tau_{cr}^\alpha$ the critical resolved shear stress for shrinkage of the twin on twin system $\alpha$. The corresponding changes in the twin volume fractions for MR and TR are thus given by:

$$\dot{f}_{MR}^\alpha = |\dot{\gamma}_{MR}^\alpha|/\gamma^{tw} \text{ and } \dot{f}_{TR}^\alpha = -|\dot{\gamma}_{TR}^\alpha|/\gamma^{tw} \quad (7)$$

where the characteristic twinning shear strain $\gamma^{tw}$ is chosen to be 0.129 for extension twinning in magnesium alloys [38].

For TP and MP, the shear rates associated with twinning system $\alpha$ can be written as:

$$\dot{\gamma}_I^\alpha = \begin{cases} -\dot{\gamma}_0 |\tau^\alpha/\tau_{cr}^\alpha|^{1/m} & \tau^\alpha < 0 \\ 0 & \tau^\alpha \geq 0 \end{cases} \quad (8)$$

where the subscript $I$ represents operations TP and MP. For TP, $\tau^\alpha$, is calculated from the stress in the twin, with $\tau_{cr}^\alpha$ the critical resolved shear stress for propagation of the twin on twin system $\alpha$. For MP, $\tau^\alpha$ is calculated from the stress in the matrix, with $\tau_{cr}^\alpha$ the critical resolved shear stress for shrinkage of the twin on twin system $\alpha$. The corresponding changes in the twin volume fraction for TP and MP are thus given by:

$$\dot{f}_{TP}^\alpha = |\dot{\gamma}_{TP}^\alpha|/\gamma^{tw} \text{ and } \dot{f}_{MP}^\alpha = -|\dot{\gamma}_{MP}^\alpha|/\gamma^{tw} \quad (9)$$

The net rate of change of the twin volume fraction associated with twinning system $\alpha$ is thus:

$$\dot{f}^\alpha = f^M(\dot{f}_{MR}^\alpha + \dot{f}_{MP}^\alpha) + f^\alpha(\dot{f}_{TP}^\alpha + \dot{f}_{TR}^\alpha) \quad (10)$$

where $f^M$ is the volume fraction of the matrix i.e. $f^M = 1 - f^{tw} = 1 - \sum_\alpha f^\alpha$.

Because it is rare for a grain to be fully twinned, a threshold for twinning in a given grain is defined as:

$$V^{th} = \min(1.0, A_1 + A_2 V^{eff}/V^{acc}) \qquad (11)$$

The combined volume fraction of all twins in a given grain is not allowed to exceed $V_{th}$ i.e. additional twinning is not allowed once $V^{th}$ has been reached, at which point the grain is said to be *twin-terminated*. $V^{acc}$ is the total accumulated twin volume fraction in the polycrystal, and $V^{eff}$ is the volume fraction of twin–terminated grains. $A_1$ and $A_2$ are optimized fitting parameters that control the evolution of $V_{th}$. For the current simulations, $(A_1, A_2)$ were assigned the values (1, 0) because the low applied strains resulted in twin volume fractions << 1.

The CRSS for twinning in a given grain is assumed to be proportional to the total volume fraction of twinning in the grain due to all twinning systems:

$$\tau = \tau_0 + (\tau_1 - \tau_0)f_{twinned}. \qquad (12)$$

while that for de-twinning is assumed to be proportional to the fraction of twinned material that has detwinned:

$$\tau = \tau_0 + (\tau_1 - \tau_0)f_{detwinned}/f_{twinned} \qquad (13)$$

In equations (12) and (13), $\tau_0$ is the CRSS for twin nucleation, and $\tau_1$ is a limiting value for the CRSS for twinning/de-twinning. The form of equation (13) is such that the CRSS for de-twinning is set to $\tau_0$ when de-twinning starts, in accordance with the observation that the CRSS for the onset of de-twinning appears to be similar or lower than the initial CRSS for twin nucleation [39].

The EVPSC-TDT model with the Affine linearization scheme is employed in the current study because the Affine scheme is considered to give the best overall performance for polycrystalline materials with face centered cubic (FCC) or HCP crystal structure [11-23].

## 4. Results and discussion

In the present work, the initial textures are discretized into 19443 orientations. The room temperature single crystal elastic constants for magnesium are taken from Simmons and Wang [40]: $C_{11} = 58$, $C_{12} = 25$, $C_{13} = 20.8$, $C_{33} = 61.2$ and $C_{44} = 16.6$ (units of GPa). The reference slip/twinning rate $\dot{\gamma}_0$ and the rate sensitivity $m$ are prescribed to be the same for all slip/twinning systems: $\dot{\gamma}_0 = 0.001 s^{-1}$ and $m = 0.05$, respectively [29,30]. Plastic deformation is assumed to occur by basal <a> slip, prismatic <a> slip, pyramidal <c+a> slip and extension twinning. The hardening parameters are obtained by fitting the initial tensile portions of the macroscopic stress strain curves for T1 and T2 (from $P^0$ to $P^1$ in Figs. 2a and 2b) and are listed in Table 1. The corresponding stress-strain response during reverse loading from $P^1$ to $P^2$ and unloading from $P^2$ are quite well predicted (Figs. 2a and 2b). The macroscopic stress-strain response for T1-CT is also quite well predicted using these model parameters(Fig. 2c).

Table 1. The hardening parameters used in the EVPSC-TDT model.

|  | $\tau_0$ | $\tau_1$ | $h_0$ | $h_1$ |
|---|---|---|---|---|
| Basal | 12 | 5 | 200 | 0 |
| Prismatic | 80 | 45 | 2000 | 0 |
| Pyramidal | 130 | 120 | 3000 | 0 |
| Extension twinning | 35 | 125 | NA | NA |

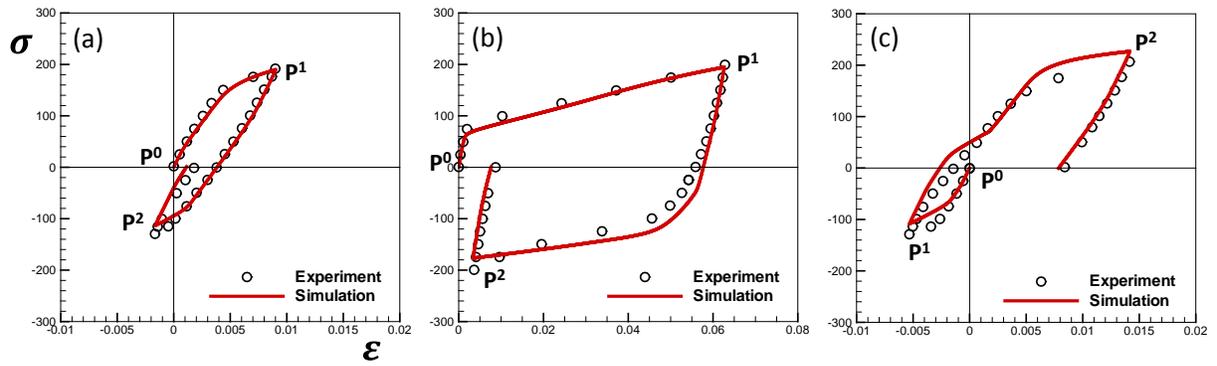

Figure 2. Experimental and simulated stress strain curves for
a) T1-TC; b) T2-TC; and c) T1-CT.

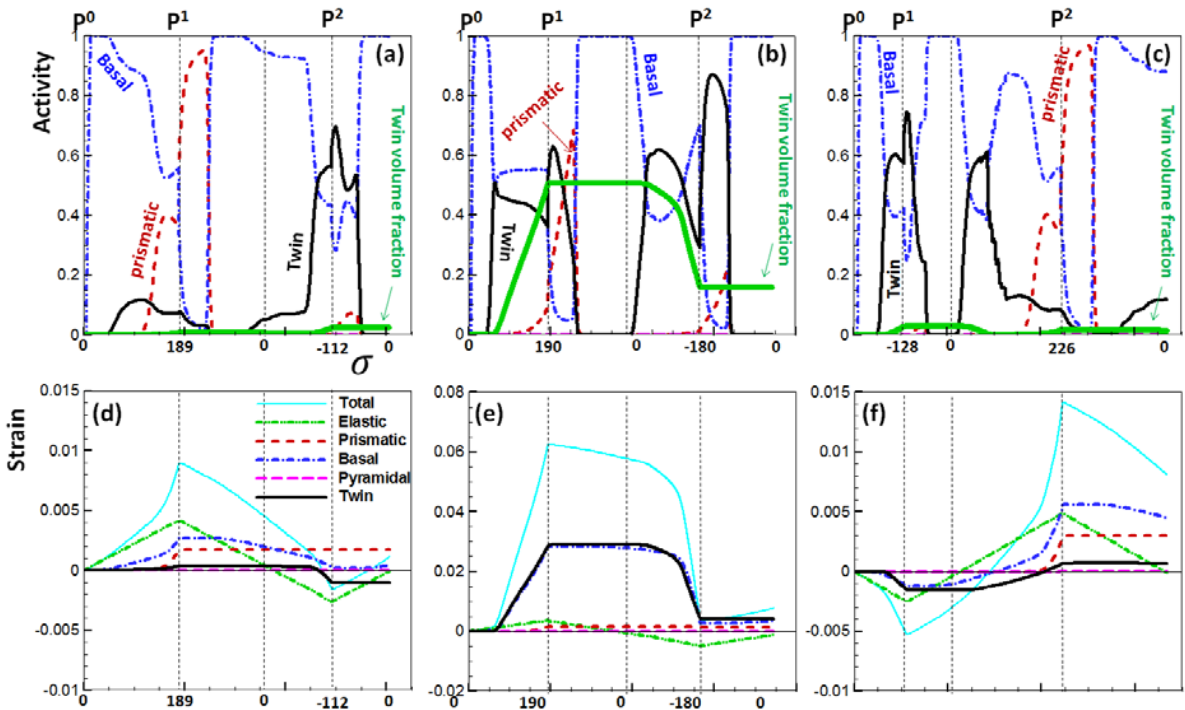

Figure 3. Predicted relative activities and contributions to the total macroscopic strain of the various deformation modes for a, d) T1-TC; b,e) T2-TC; and c, f) T1-CT.

In order to investigate the contribution of various deformation mechanisms especially when the absolute value is small, the relative activity is defined as the fraction of the total shear rate of a deformation mechanism to the total shear rate of all deformation mechanisms. The predicted relative activities and the contributions to the overall macroscopic strain of the various deformation mechanisms for the three tests are presented in Fig. 3. The relative activity of a given deformation mode is the ratio of the corresponding plastic shear rate to the total plastic shear rate due to all the deformation modes at the applied stress.

For T1-TC (Figs. 3a,d), basal slip starts almost immediately during forward tensile loading and is the only active system until extension twinning is activated at ~50 MPa. For texture T1, most grains are unfavorably oriented for extension twinning, which severely limits the volume fraction of twinning. As a result, extension twinning accounts for no more than ~10% of the total shear rate at any point during forward tensile loading (Fig. 3a), and contributes very little to the macroscopic strain (Fig. 3d). Prismatic slip is first activated at ~120 MPa, but does not contribute significantly to the macroscopic strain until ~150-170 MPa, which corresponds to general yielding in the macroscopic stress-strain curve. At this stress, there is also an increase in the rate at which basal slip contributes to the macroscopic strain, from which it can be concluded that general yielding in forward tension for texture T1 occurs via a combination of basal and prismatic slip, with twinning contributing very little. In the early stages of unloading from forward tension, from the maximum tensile stress to ~150 MPa, prismatic slip, basal slip and twinning are all active, but contribute negligibly to reverse plastic deformation. Plastic deformation in the latter stages of unloading and the early stages of reverse loading in compression are dominated by basal slip, with negligible contributions from prismatic slip and twinning, indicating that basal slip is responsible for the non-linear unloading behavior in the macroscopic stress-strain curve. Fig. 3d shows that yielding in reverse compression coincides with a rapid increase in twinning activity, which contributes most of the macroscopic plastic strain, as well as a slight increase in the rate at which basal slip contributes to plastic strain. Though the relative activity of twinning during reverse loading in compression is high, particularly between the yield point and $P^2$, the twin volume fraction remains small (~3%) because of the low applied strain. Fig. 2a shows that the model underpredicts the hardening behavior in compression. Similarly to unloading from forward tension, all three deformation modes are active but contribute negligibly to macroscopic plastic deformation in the early stages of unloading from compression, and the non-linear behavior is due to mainly to basal slip.

For T2-TC (Figs. 3b,e), basal slip again starts almost immediately during forward tensile loading and is responsible for almost all of the plastic deformation until macroscopic yielding at

~50-60 MPa, which corresponds to a sharp increase in twinning activity. From the yield point up to the maximum tensile stress, twinning and basal slip contribute almost equally to the macroscopic plastic strain. The large contribution of twinning is expected because most of the grains are oriented favorably in texture T2. The large contribution of basal slip is also associated with twinning, since the re-orientation due to twinning results in a marked increase in the Schmid factor for basal slip in the twinned domains. This increase in the Schmid factor has two related effects: 1) the resolved shear stress for basal slip is increased, making it easier to activate basal slip, and 2) the macroscopic strain increment per unit shear on the basal system is also increased. Unlike T1-TC, the onset of prismatic slip occurs well after general macroscopic yielding, though the corresponding applied stress is similar. As shown in Fig. 3e, however, prismatic slip contributes very little to the macroscopic strain. As for T1-TC, twinning and prismatic slip are stable during unloading and in the early stages of reverse loading. The model predicts that the limited nonlinearity in the macroscopic stress-strain curve over this stress interval is due mainly to basal slip. However, Fig. 2b shows that the model significantly underestimates the degree of non-linearity in this interval. Yielding in reverse compressive loading is due almost equally to detwinning and basal slip. The model predicts significant twinning activity during both forward loading and reverse loading, with the total twin volume fraction increasing steadily during forward tensile loading up to ~50%, remaining constant during unloading and in the early stages of reverse compressive loading, then decreasing in two distinct stages during further reverse compressive loading. The model predicts that the limited non-linearity in the stress-strain during the final unloading step is due mainly to basal slip.

For T1-CT (Figs. 3c,f), basal slip again starts almost immediately during forward compressive loading and is responsible for almost all of the plastic deformation until macroscopic yielding at ~50-60 MPa, which corresponds to a sharp increase in twinning activity. From the yield point up to the maximum compressive stress, twinning and basal slip contribute almost equally to the macroscopic plastic strain, as in T2-TC. The large contribution of twinning is expected because most of the grains are oriented favorably in texture T1. Also as in T2-TC, the large contribution of basal slip is associated with lattice re-orientation due to twinning. Prismatic slip contributes negligibly during the twin-dominated initial compression step. Fig. 2c shows that the model significantly underestimates the hardening behavior during this step. As in the previous two tests, the model predicts that twinning is stable during unloading and in the early stages of reverse loading such that the limited nonlinearity in the macroscopic stress-strain curve over this stress interval is due mainly to basal slip. Fig. 2c shows that the model predicts a much more pronounced inflection point in the macroscopic stress-strain curve during unloading from compression and reloading in tension than is actually observed. Furthermore, the model significantly underpredicts the amount of plastic deformation during unloading. Yielding in reverse tensile loading, as in forward tensile loading in T1-TC, is due to a combination of basal slip, prismatic slip, and twinning. Finally, as in all of the unloading steps in all of the tests, the observed non-linearity is due almost entirely to basal slip. Fig. 3c shows that the twin volume fraction increases during forward compressive loading (up to ~5%), stays constant during unloading, and increases again during reverse loading.

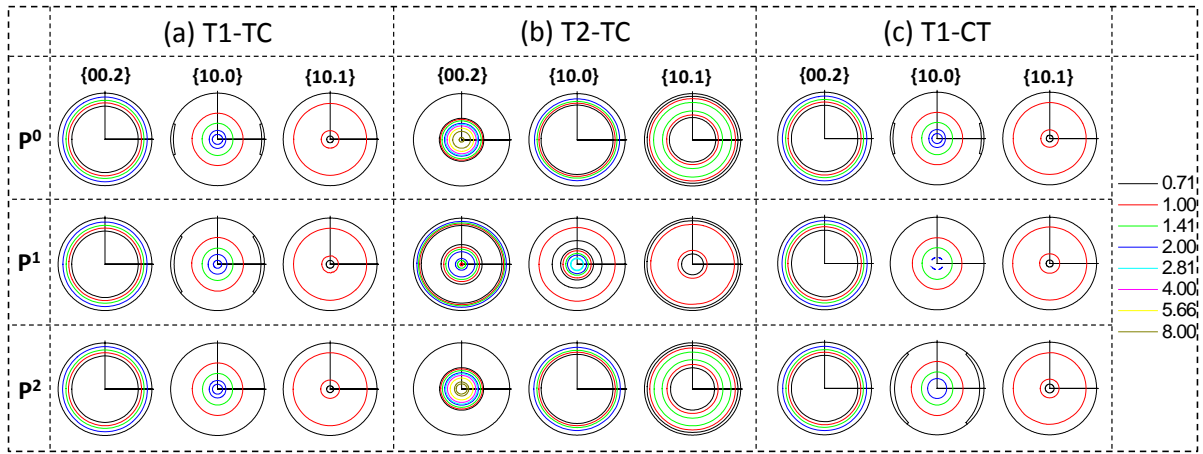

Figure 4. Predicted textures at $P^1$ and $P^2$ for (a)T1-TC, (b) T2-CT and (c) T1-CT.

Fig. 4 shows the predicted textures at $P^1$ and $P^2$ for the three tests. The evolution of the texture is small for T1-TC and T1-CT, mainly because of the relatively low applied strains during the compression stages, for which most of the grains are favorably oriented. In contrast, the evolution of the texture is significant for T2-TC because the majority of grains are favorably oriented for twinning in tension, for which the applied strain is quite large. Twinning results in a marked texture evolution because it results in a reorientation of the parent lattice by 86.3° (from the center of {00.2} pole figure at $P^0$ to the outer ring at $P^1$), while de-twinning reorients the twinned grains back.

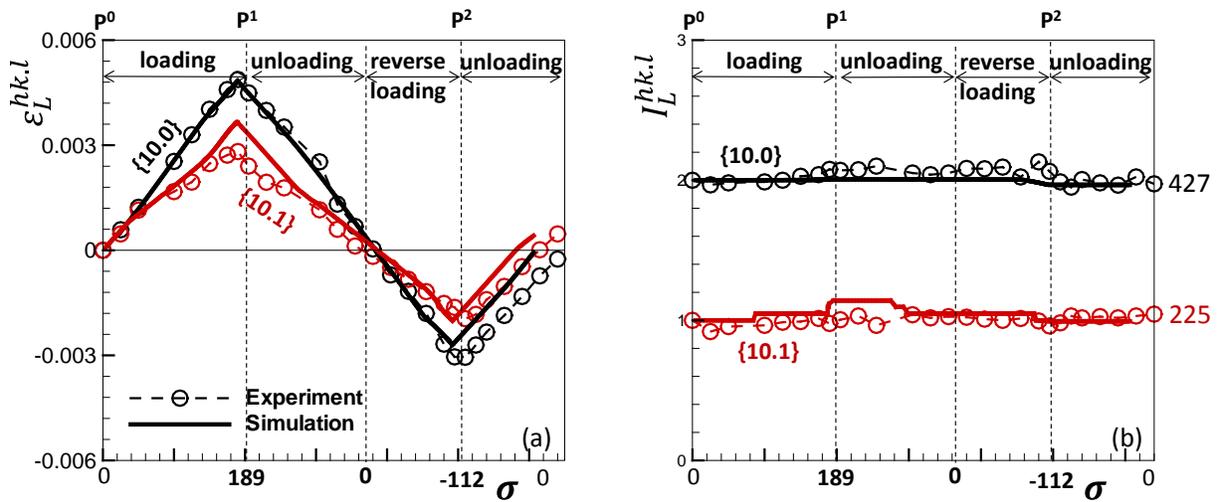

Figure 5. The evolution of (a) internal elastic strain $\varepsilon_L^{hk.l}$ and (b) normalized intensity $I_L^{hk.l}$ of T1-TC.

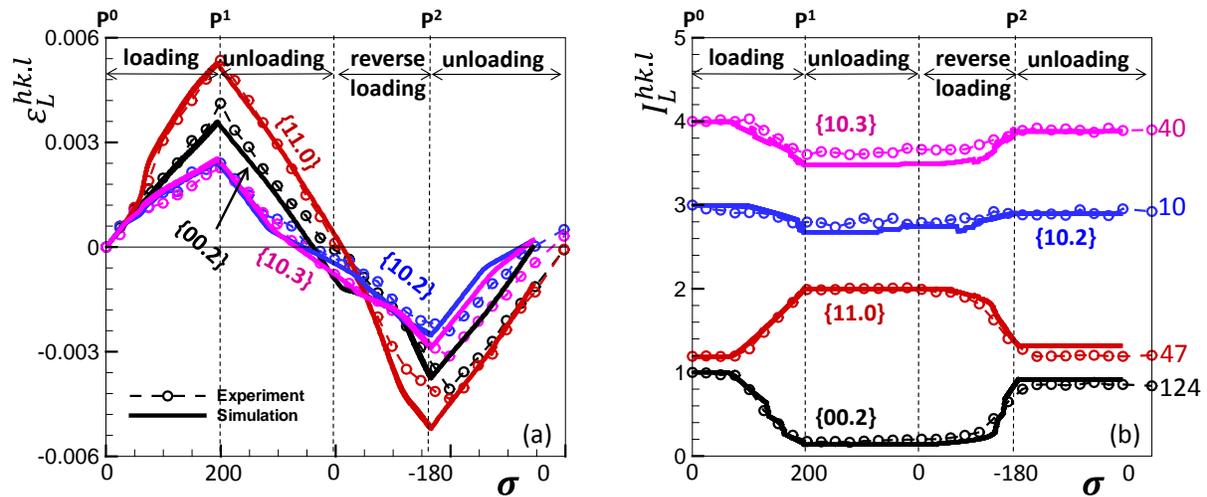

Figure 6. The evolution of (a) internal elastic strain $\varepsilon_L^{hk.l}$ and (b) normalized intensity $I_L^{hk.l}$ of T2-TC.

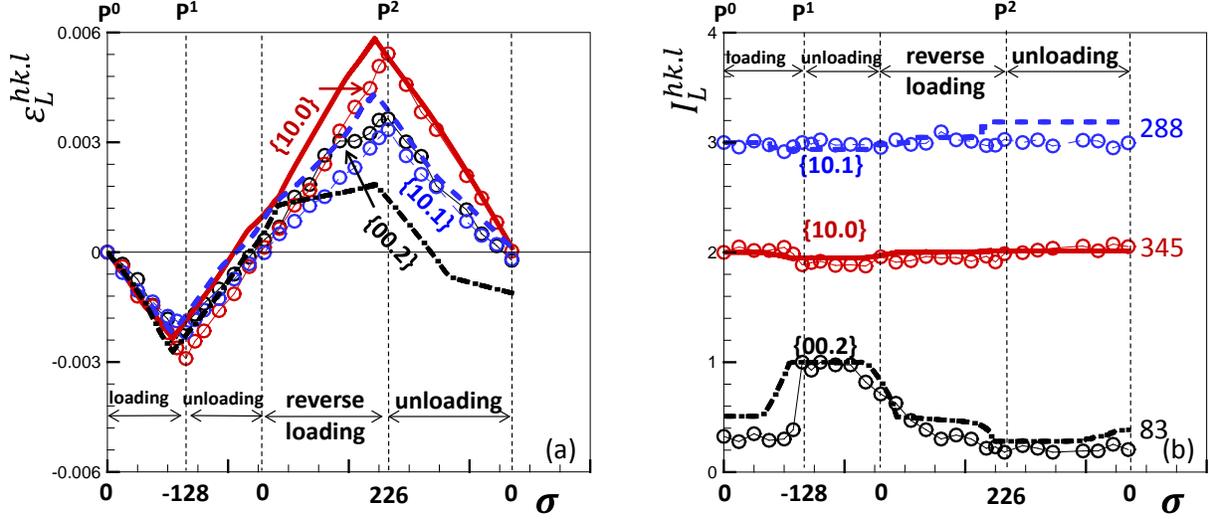

Figure 7. The evolution of (a) internal elastic strain $\varepsilon_L^{hk.l}$ and (b) normalized intensity $I_L^{hk.l}$ of T1-CT.

The measured and predicted elastic lattice strains ($\varepsilon_L^{hk.l}$) are shown in Figs 5a, 6a, and 7a for T1-TC, T2-TC, and T1-CT, respectively, while the corresponding measured and predicted diffraction peak intensities ($I_L^{hk.l}$) are shown in Figs. 5b, 6b, and 7b. In these figures, the lattice strains and intensities are associated with different grain families, identified by the crystallographic plane normal to the axis of applied stress. Thus the {10.0} data correspond to a grain family in which the {10.0} plane is normal to the stress axis. The absolute diffraction intensities depend on many factors including the volume fraction of grains involved, the scattering angle and the texture of the sample. As a consequence, it is difficult to calculate absolute intensities for comparison with the experiment. Instead, a relative diffraction intensity, defined as the instantaneous intensity normalized by the corresponding initial intensity (the maximum intensity for {10.0} is used to minimize the uncertainty from the low intensity), is used here. The absolute values used to normalize the experimental intensities are included in the figures (next to each curve). To avoid congestion, the original relative intensities are shifted by one unit parallel to ordinate.

For T1-TC (Fig. 5), the {10.0} lattice strains during tensile loading and unloading are well predicted by the model; however, the {10.1} lattice strains are not as well predicted. The {10.0} grain family bears more stress because it is unfavorably oriented for basal slip and extension twinning. As a result, the lattice strains increase linearly with applied stress, until prismatic slip can be activated at higher stresses. The {10.1} grain family is favorably oriented for basal slip, resulting in lower lattice strains during loading in tension. The lattice strains for both grain families during reverse compression are well simulated by the model, except that the lattice strains for the highest compressive stress are significantly underpredicted. As mentioned earlier, this issue is due to a failure of the model to adequately capture the hardening in compression when twinning activity is significant. The normalized diffraction intensities vary little during the test because the applied strain is small and therefore twinning activity is limited. However, close inspection reveals that the small increase in $I_{\{10.0\}}$ that occurs during forward tension is not captured by the model, while the model does predict the small decrease that occurs during compression. The model also predicts small changes in $I_{\{10.1\}}$ that do not appear in the experimental data.

For T2-TC, the measured and predicted lattice strains for the {00.2}, {11.0}, {10.2} and {10.3} grain families are shown in Fig. 6a. The figure shows that the lattice strains for forward tension are very well predicted by the model, with the exception of the {00.2} lattice strains, which are well predicted up to the penultimate loading step, after which the model underpredicts the lattice strain. The {11.0} grain family is hard in tension because it is unfavorably oriented for basal slip and extension twinning. The {10.2} and {10.3} grain families are favorably oriented for basal slip and therefore yield easily. The {00.2} grain family is poorly oriented for basal slip but is favorably oriented for extension twinning, such that its behavior in tension is intermediate between those of the {11.0} and {10.2}/{10.3} grain families. In reverse compression ($P^1$ to $P^2$), the model captures the lattice strain evolution fairly well though the response of the {11.0} family is significantly overpredicted. The evolution of the diffraction intensities for the {00.2} {11.0}, and {10.3} grain families, which can be attributed to extension twinning over the small strain intervals studied, is captured very well by the model. The decreased intensity of the {00.2} grain family during forward loading indicates that these grains undergo extension twinning, which is associated with a reorientation of the c-axis by 86.3° and making it perpendicular to the ED. Simultaneously the intensity of the {11.0} grain family increases because some grains are twinned and reoriented into this orientation. During reverse loading ($P^1$ to $P^2$), the intensity of the {00.2} grain family increases and that of the {11.0} grain family decreases due to twinning and detwinning. The intensity of the {10.3} grain family varies similarly to that of the {00.2} family. The intensity of {10.2} decreases during forward loading, stays constant during unloading and the early stages of reverse loading, increases again during the later stages of reverse loading, then stays constant during the final unloading step. The decrease in intensity during forward loading indicates that grains in this family undergo twinning, as expected.

Magnesium alloys with texture T2, in which extensive extension twinning can be activated by tension parallel to the extrusion direction (ED), have not been studied by in-situ neutron diffraction previously [15-17]. Conventional extruded samples have the typical basal texture (e.g., Fig. 1a), on which deformation twinning is activated mainly by compression along the ED (compression perpendicular to c-axis). Conventional rolled plate also has a typical basal texture, which can be activated by either compression perpendicular to normal direction (ND) of the plate (compression perpendicular to the c-axis) or tension along ND (tension along the c-axis). A requirement for in-situ neutron diffraction experiments is that the sample have a sufficiently long gauge length, which often precludes the possibility of performing measurements on samples having the loading axis along ND.

The measured and predicted internal elastic strains for the {00.2}, {10.0}, and {10.1} grain families for T1-CT are compared in Fig. 7a. The lattice strain evolution in all of the grain families is similar in forward compression. The model significantly overpredicts the behavior of the {10.0} and {10.1} grain families in reverse tension, while greatly underpredicting the behavior of the {00.2} family. This combination of discrepancies between the model and the measurements at the grain level nevertheless leads to an acceptable fit of the macroscopic stress-strain curve, demonstrating the great value of neutron diffraction data in validating models of polycrystal plasticity. The predicted diffraction intensities are in fairly good agreement with the experimental data (Fig. 7b). The {00.2} intensity increases during forward loading ($P^0$ to $P^1$), while the {10.0} intensity decreases concurrently. When the sample is unloaded after compression (from $P^1$), the intensity of the {00.2} peak remains stable down to an applied stress of ~–50 MPa, and then decreases gradually. The intensity falls by about 40% during unloading to zero load, indicating that about 40% of the twinned volume has de-twinned during unloading after compression. During reverse loading in tension, de-twinning continues until the {00.2} intensity at the start of the test is fully recovered at about 100 MPa. The {00.2} intensity continues to decrease at a lower rate with increasing load from 100MPa to 226MPa. At this point, it appears that the {00.2} minority grains undergo {10.2} extension twinning. During unloading after the tensile portion of the stress-strain curve, the {00.2} intensity does not change, suggesting that the twinned material in the minority {00.2} grains does not undergo significant de-twinning.

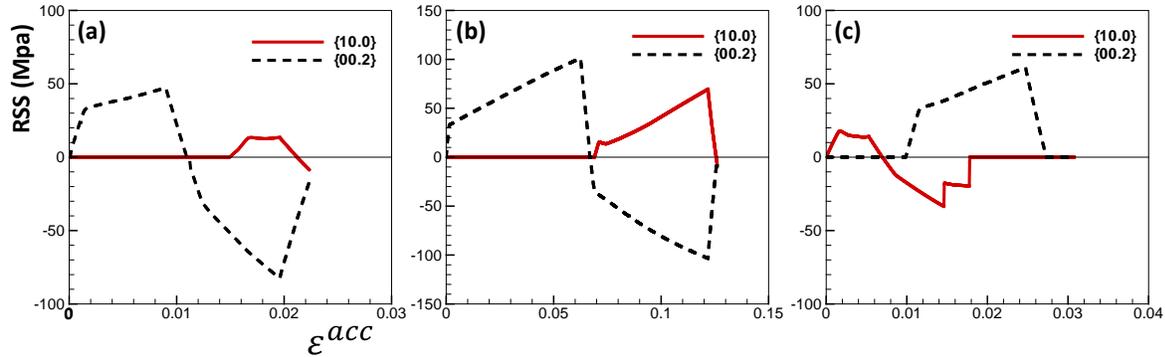

Figure 8. Effective resolved shear stress of twinning systems in {10.0} and {00.2} grain families under (a) T1-TC, (b) T2-TC, and (c) T1-CT.

The effective resolved shear stresses of twinning system in {10.0} and {00.2} initial grain families are shown in Figure 8. The effective resolved shear stress is defined as the weight-averaged resolved shear stress of active twinning systems over the grains in the family. When a new grain associated with a twinning system is created, the corresponding RSS of the twinning system in the new grain is accounted for in calculating the effective RSS of its parent grain. The higher the effective RSS (>0) is, the easier the grain family to twin. Upon loading reversal, the higher the magnitude of RSS (<0) is, the easier the grain family to detwin. As can be seen, the {00.2} grain family in both T1 and T2 samples favors twinning under tension, and detwinning under reverse compression. On the contrary, the {10.0} grain family favors twinning under compression, and detwinning under reverse tension. This is consistent to our experimental observations.

The evolution of the diffracted intensity reflects the volume fraction change due to twinning. The good agreement between experiment and prediction reflects that twinning (and detwinning) is largely "kenematically driven", which means the twin volume fraction is largely dependent on the applied strain, not stress. However, the internal elastic strain reflects the stress distribution among the grains, which is affected by the interaction between parent grain and twin and interaction among all grains. To improve the predictability in future work, TDT model should consider the relevant issues including stress relaxation due to twin initiation, back stress due to the twinning induced plastic deformation, etc.

## 5. Concluding remarks

The macroscopic stress-strain response, the texture evolution, and the evolution of internal elastic strain of extruded Mg-8.5wt.%Al alloys under cyclic deformation are systematically investigated. The following conclusions can be made:

1. Both in-situ neutron diffraction and EVPSC-TDT model are employed to investigate the behavior of Mg-8.5wt.%Al alloys in terms of macroscopic stress strain curves, activities of deformation mechanisms, evolution of textures, evolution of internal elastic lattice strain and diffraction peak intensities.

2. Extension twin is usually activated by compression perpendicular to the c-axis of grain in conventional in-situ neutron diffraction measurements. The prepared sample with a modified initial texture (T2) successfully activates extension twinning by tension along the c-axis of the grains, providing a unique behavior of magnesium alloy.

3. The internal elastic strains and the diffraction intensities are measured by the *in-situ* neutron diffraction and are simulated by the EVPSC-TDT model. The results show twinning and/or de-twinning plays important roles during cyclic deformations for the alloys with different initial textures. The twinning and de-twinning change the orientation of the grains significantly therefore evolve the intensities of the diffraction peaks. The good agreement between the predictions and the experiments validates that the EVPSC-TDT model can well capture the twinning and de-twinning activities frequently observed during deformation of magnesium alloys.

**Acknowledgements**

HW and PDW were supported by the Natural Sciences and Engineering Research Council of Canada (NSERC) and the Ontario Ministry of Research and Innovation (OMRI). SYL would like to thank the support from the National Research Foundation of Korea (NRF) grant funded by the Korean government (MSIP) (Nos. 2013R1A4A1069528 and 2013R1A1A1076023).